\documentclass[10pt, conference, letterpaper]{IEEEtran}

\usepackage{cuted}
\usepackage{caption}
\usepackage{lipsum}
\usepackage{indentfirst}
\usepackage{booktabs}

\usepackage{cite}

\usepackage[pdftex]{graphicx}
\usepackage{wrapfig}
\usepackage{subfig}
\usepackage{float}

\usepackage{amsfonts, eqnarray, amsmath, amssymb}
\usepackage{mathptmx}
\usepackage{mathtools}
\usepackage{xfrac}
\usepackage[cmintegrals]{newtxmath}
\usepackage{breqn}

\usepackage{algorithmic}

\usepackage{array}

\begin{document}
%
\title{CCOMPASSION: A Hybrid Cloudlet Placement Framework over Passive Optical Access Networks}

\author{\IEEEauthorblockN{Sourav Mondal$^1$, Goutam Das$^2$ and Elaine Wong$^1$}
\IEEEauthorblockA{$^1$Dept of Electrical and Electronic Engineering, The University of Melbourne, VIC 3010, Australia.\\
$^2$G. S. Sanyal School of Telecommunications, Indian Institute of Technology Kharagpur, 721302, India.}}

\maketitle

\begin{abstract}
Cloud-based computing technology is one of the most significant technical advents of the last decade and extension of this facility towards access networks by aggregation of cloudlets is a step further.  To fulfill the ravenous demand for computational resources entangled with the stringent latency requirements of computationally-heavy applications related to augmented reality, cognitive assistance and context-aware computation, installation of cloudlets near the access segment is a very promising solution because of its support for wide geographical network distribution, low latency, mobility and heterogeneity.  In this paper, we propose a novel framework, Cloudlet Cost OptiMization over PASSIve Optical Network (CCOMPASSION), and formulate a nonlinear mixed-integer program to identify optimal cloudlet placement locations such that installation cost is minimized whilst meeting the capacity and latency constraints.  Considering urban, suburban and rural scenarios as commonly-used network deployment models, we investigate the feasibility of the proposed model over them and provide guidance on the overall cloudlet facility installation over optical access network.  We also study the percentage of incremental energy budget in the presence of cloudlets of the existing network.  The final results from our proposed model can be considered as fundamental cornerstones for network planning with hybrid cloudlet network architectures.
\end{abstract}

\section{Introduction}\label{sec1}
Riding on the shoulders of the gigantic technical advancements during the past few decades, the vision of \emph{ubiquitous computing in the 21st century} published by Weiser in his seminal paper \cite{weiserm1991}, has managed to land at a solid ground of reality of present times.  In smart cities and smart homes, almost every object around us is expected to have some computational capability and connectivity to the Internet.  According to the prediction from Gartner, when Internet of Things (IoT) becomes completely commercialized along with the standardization of 5G technologies, nearly 20 billion of devices will connect to the Internet by 2020 \cite{Gartner}.  In conjunction with this, mobile data traffic has already increased by 63\% in 2016, and by 2021, this is expected to increase by seven folds further \cite{Cisco}.\par
The technology that provides distributed computation facilities on battery-powered, portable and mobile devices which are interconnected via mobile communication standards and protocols, is known as \emph{Mobile Computing} \cite{mobile_wiki}.  However, along with several essential advantages, there exist some crucial limitations in mobile computing, e.g., resource constraints on battery, degradation of Quality-of-Service (QoS), limited bandwidth and connection latency.  Some typical smart applications related to augmented reality, autonomous transport, and cognitive assistance, demand a network latency of 1-10 ms \cite{Ha}.  Moreover, mobile devices are designed to be lightweight and portable, which makes them poor in computational resources and memory.  Therefore, most often meeting both the latency requirements as well extended battery life become a trivial challenge for the mobile devices and offloading the major computational tasks to cloud servers appeared to be highly essential \cite{Kumar}.\par
\begin{figure}[t!]
\centering
\includegraphics[width=0.5\textwidth,height=4.5cm]{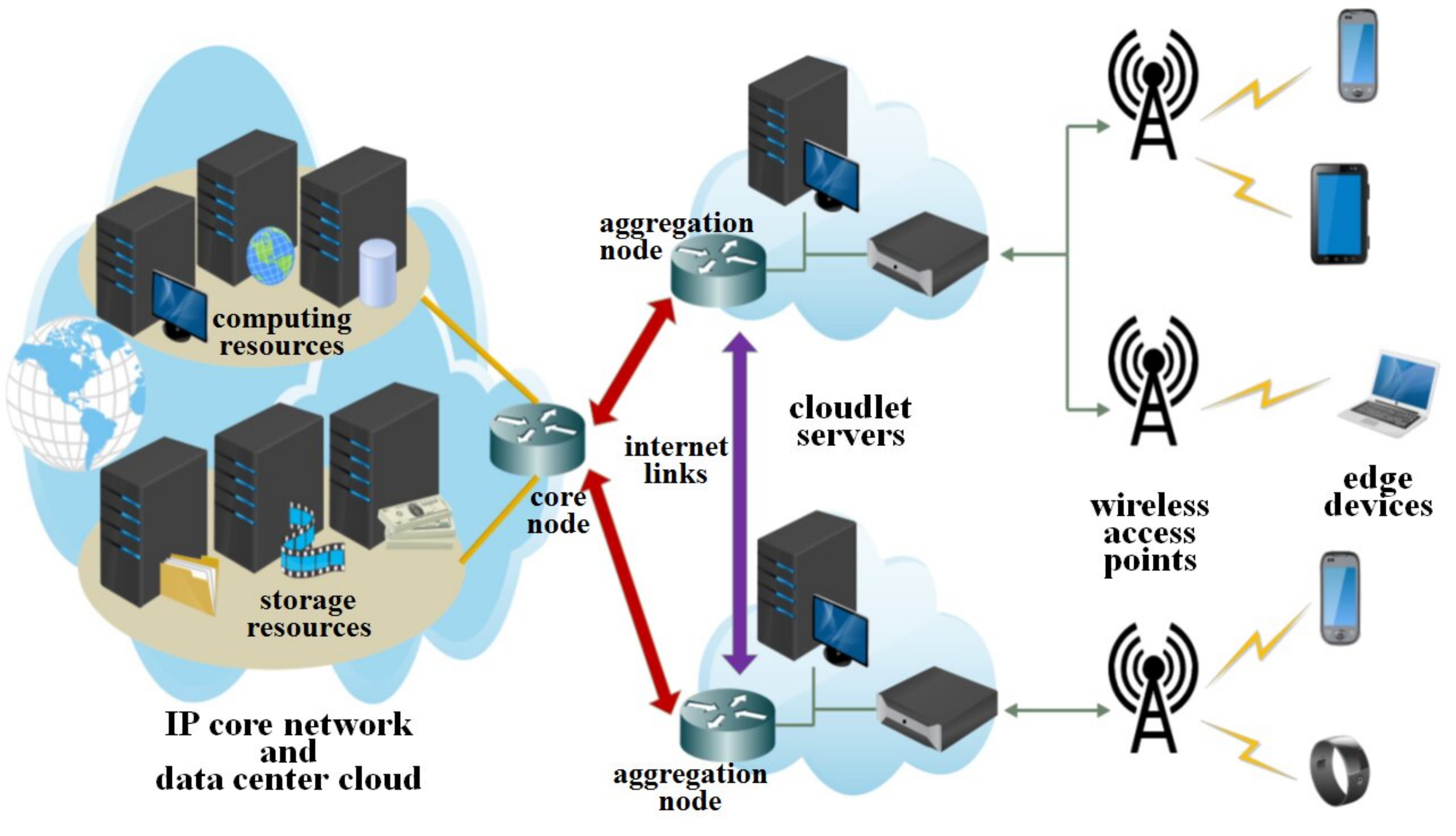}
\caption{A simplified top level architecture of cloud-cloudlet-edge devices network infrastructure \cite{Satyanarayanan}.}
\label{cloud_cloudlet}
\end{figure}
\setlength{\textfloatsep}{5pt}%
The process of offloading the intensive computation to the cloud server by the mobile devices is known as \emph{cyber foraging} \cite{Balan}.  However, cloud servers are installed at geographically secure locations, beyond the core backbone network, and hence the cloud server access should not be taken for granted.  In disaster recovery scenarios or secure military operation areas, where basic network infrastructures have been disrupted, remote cloud access may become impossible \cite{Satya2}.  Due to multi-hop network connectivity, not only the end-to-end network latency becomes large, as pointed out by authors in \cite{Kumar}, maintenance of privacy, security and reliability of the offloaded data by users becomes challenging too.  Nonetheless, when the end-to-end latency is less than QoS latency requirement, the best decision for the mobile devices is to offload \cite{Kumar}.\par
To mitigate these aforementioned challenges, the authors of \cite{Satyanarayanan} proposed the idea of \emph{cloudlets} as a small-scale but distributed cloud computation facility at close proximity of end-users in the access segment.  A typical cloudlet-empowered cloud computation network consist of a \emph{three-tier architecture}, as shown in Fig. \ref{cloud_cloudlet}.\par
The authors of \cite{close} made a comparative study to understand the impact of distance on end-user experience especially with transient display applications.  Both remote cloud and single-hop cloudlets were used to offload computational tasks in their experiment.  Through their experimental results, the authors of \cite{Ha} confirmed that execution of the offloaded task by a nearby computer is always more latency efficient over a distant server.  The authors of \cite{Grace} further demonstrated that by adopting efficient virtual machine (VM) provisioning algorithms, the computation latency can be reduced to a great extent for tactile applications.\par
Installation of cloudlet facilities at suitable locations is becoming a primary design challenge in satisfying the QoS latency requirements of latency-sensitive applications.  The authors of \cite{Ceselli} presented link-path mixed integer linear programming (MILP) formulations for \emph{static planning} as well as \emph{dynamic planning} with VM bulk migration and live migration over a 4G cellular network.  The authors of \cite{efficient} formulated an Integer Linear Programming (ILP) problem for cloudlet placement to minimize the \emph{cloudlet access delay} in a Wireless Metropolitan Area Network (WMAN) with several wireless Access Points (WAPs) and proposed an efficient heuristic algorithm for faster convergence of their optimization model.  On a similar note, the authors of \cite{MJia} proposed a heuristic algorithm to minimize the \emph{system response time} by optimally allocating cloudlets to end-users in a WMAN.  Moreover, the authors of \cite{Ansari} proposed an architecture that minimizes the carbon footprint of cloudlet network by adopting harnessing and usage of green energy.  Note that the authors in these works had considered only wireless access media.\par
On the contrary, the authors of \cite{opt_cloud} highlighted several benefits of using \emph{optical access networks} for cloud computing applications due to its cost-effectiveness, high-bandwidth data transmission, efficient network virtualization and network scalability.  Moreover, the authors of \cite{Martin} outlined a novel cloud and cloudlet empowered fiber-wireless (FiWi)-heterogeneous network architecture for LTE-Advanced (LTE-A).  The authors of \cite{Martin2} proposed a cloudlet-aware resource management scheme to reduce offload delay and prolong mobile-devices' battery life by incorporating offloading activities into the underlying FiWi dynamic bandwidth allocation process.  In \cite{Imali}, the authors implemented a cloudlet framework with control server at the central office (CO) of a fiber-based access network for human-machine interactive applications.\par
However, to the best of our knowledge, the optimized design and planning of \emph{hybrid placement of cloudlets based on TDM-PON support architecture} is an unexplored area.  Optical access network standards like 10G(E)-PON has been the preferred technological solution of broadband access due to its low cost per bit and very high bandwidth support.  Moreover, the evolution of radio-over-fiber (RoF) and FiWi makes the Internet accessible to a huge number of end users \cite{Elaine}.  This motivates us to propose a hybrid cloudlet support architecture based on an existing 10G(E)-PON infrastructure and develop a framework that minimizes the capital expenditure (CAPEX), mainly subjected to the cloudlet computational capacity and latency constraints.\par
In this paper, we propose the end-to-end hybrid cloudlet placement architecture of Cloudlet Cost OptiMization over PASSIve Optical Networks (CCOMPASSION), with an existing tree-and-branch TDM-PON based network infrastructure.  The cloudlets can be suitably placed in the field, remote node (RN) and optical line terminal (OLT) or CO locations depending upon computation task requirements.  We formulate a mixed-integer non-linear programming (MINLP) problem to identify cost efficient cloudlet placement locations subjected to optimized connectivity with optical network units (ONUs).  In this work, our main focus is on \emph{static planning} of the network.  Our model identifies optimal cloudlet placement locations as well as optimal computational resources, i.e., number of racks in each cloudlet.  This helps to prevent under or over-utilization of the resources.  We further \emph{linearize the objective function} so that our design tool can compute the optimal solution with a much faster convergence rate.  Our primary design constraint of the optimization problem is to \emph{achieve a very low end-to-end latency} ideal for tactile applications, i.e., 1 ms, 10 ms and 100 ms.  We validate our model against urban, suburban and rural scenarios and present the comparative deployment cost, workload distribution among field, RN and CO cloudlets, average number of racks per cloudlet and the increase of energy budget over a typical 10G(E)-PON, against the targeted latency figures of 1 ms, 10 ms and 100 ms.\par
The rest of this paper is organized as follows.  In Section \ref{sec2}, the details of TDM-PON based hybrid cloudlet placement network architecture are described.  In Section \ref{sec3}, the system model and the MINLP formulation are presented.  The simulation results are presented for discussion in Section \ref{sec4}.  Finally, Section \ref{sec5} summarizes the key observations of the entire work.\par
\begin{figure*}[t!]
\centering
\includegraphics[width=\textwidth]{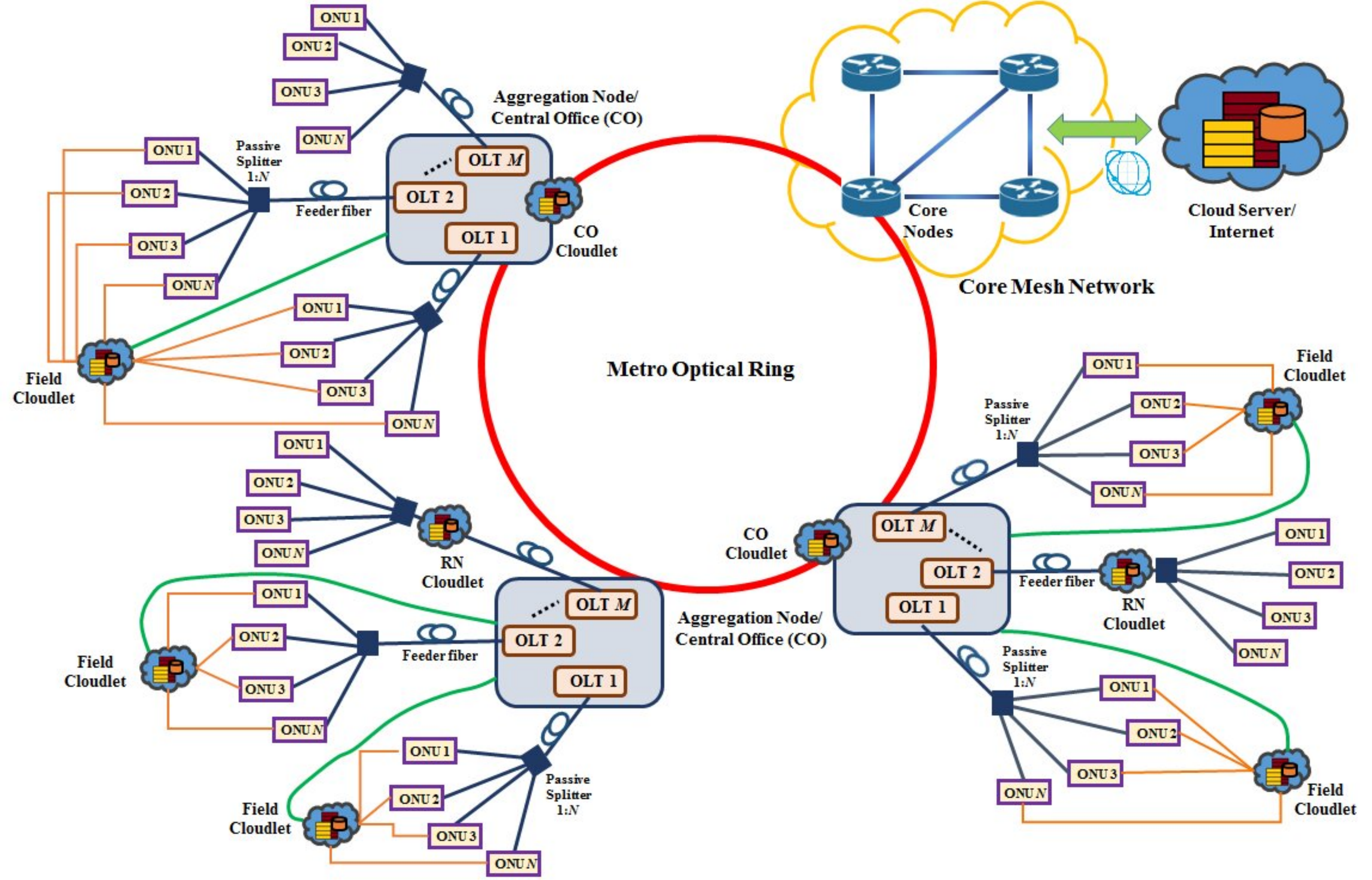}
\caption{Complete physical network infrastructure of the proposed CCOMPASSION framework showing cloudlet placement locations in the field, remote node and central office.}
\label{hybrid_cloudlet}
\end{figure*}
\setlength{\textfloatsep}{5pt}%

\section{The CCOMPASSION Framework}\label{sec2}
The proposed end-to-end physical network infrastructure of hybrid cloudlet support network based on TDM-PON access, i.e., Cloudlet Cost OptiMization over PASSIve Optical Network (CCOMPASSION) is illustrated in Fig. \ref{hybrid_cloudlet}.  With a conventional wireless access network, the base stations (BSs) or WAPs form a star connection with the aggregation node or CO \cite{Ceselli}.  Nevertheless, with optical access networks, the connection among the ONUs and cloudlets in the field, RN or CO need not necessarily follow a star topology.  With a tree-and-branch network topology, as shown in Fig. \ref{hybrid_cloudlet}, the split ratio of the passive splitter installed at RN is chosen according to number of end-users and their bandwidth requirements, e.g., 1:4 or 1:8 or 1:16 \cite{Elaine}.\par
In a hybrid cloudlet placement scheme, cloudlets can be installed either in the field, RN or CO locations, so that close proximity with the end-user is maintained.  Additionally, WAPs with WiFi or LTE/5G support can be integrated with ONUs to provide network coverage to a larger number of end-users.  A typical cloudlet rack is expected to support around 2500 VMs and with a higher the number of end-user connectivity, higher number of racks are required \cite{Ceselli}.  The cloud servers are in general placed at a geographically secured location, beyond backhaul metro networks, as shown in Fig. \ref{hybrid_cloudlet}.\par
In the CCOMPASSION framework, for the \textit{field cloudlets}, point-to-point fiber links (with datarate 1 Gbps) between a cloudlet and each ONU connected to it (brown links) are installed.  An additional point-to-point fiber link between each cloudlet and the nearest CO (green links) is also installed to create the cloudlet-cloud connectivity.\par
However, the idea of \emph{installing cloudlets at RN locations} is more beneficial over field cloudlets, especially in dense deployment scenarios, due to the fact that installation of a field cloudlet carries additional point-to-point new fiber link installation costs, whereas installing a cloudlet at the RN allows us to re-use the existing fiber links between ONUs and RNs.  We can use just one or more additional wavelengths to be time-shared amongst the ONUs for communicating with the cloudlet placed at the RN.\par
Interestingly, \emph{installing a cloudlet at the CO} is the most economical option amongst all three as the infrastructure setup cost is lowest, but the key drawback arises in terms of transmission latency and bandwidth.  The COs are geographically farther from the ONUs, field and RN locations, and the ONUs use the default uplink and downlink channels for communication with cloudlets placed at the CO.  In general, the average background load for downlink and uplink can be considered to be around 70\% and 50\%, respectively \cite{Machuka} and hence, the ONUs can use the unused bandwidth for communication with cloudlet placed at CO.  In this case, the transmission latency can potentially be higher under high load conditions.\par
\begin{table}[]
\centering
\caption{Hybrid Cloudlet Network Optimization Parameters}
\label{table1}
\resizebox{\columnwidth}{!}{%
\begin{tabular}{cl}
\hline
\multicolumn{1}{l}{\textit{\textbf{Symbol}}} & \multicolumn{1}{c}{\textit{\textbf{Definition}}}                                                                                                             \\ \hline
$A$                                          & Set of possible field cloudlet locations                                                                                                                     \\
$B$                                          & Set of possible RN cloudlet locations                                                                                                                        \\
$C$                                          & Set of possible CO/OLT cloudlet locations                                                                                                                        \\
$D$                                          & Set of locations of existing ONUs in the TDM-PON network                                                                                                     \\
$K$                                          & Set of number of racks in a cloudlet location {[}1,10{]}                                                                                                     \\
$\alpha$                                     & Cost of installing a single rack cloudlet at field, RN or CO                                                                                                 \\
$\xi_a$                                      & New infrastructure installation cost at field location $a\in A$                                                                                              \\
$\xi_b$                                      & New infrastructure installation cost at RN location $b\in B$                                                                                                 \\
$\xi_c$                                      & New infrastructure installation cost at CO location $c\in C$                                                                                                 \\
$\eta$                                       & Cost of installing new optical fiber per kilometer                                                                                                           \\
$L_{max}$                                    & \begin{tabular}[c]{@{}l@{}}Maximum allowed fiber length between field cloudlet \\ $a\in A$ and ONU $d\in D$\end{tabular}                                     \\
$L_{ad}$                                     & \begin{tabular}[c]{@{}l@{}}Actual optical fiber length between field cloudlet $a\in A$ \\ and ONU $d\in D$\end{tabular}                                      \\
$x_{bd}^{adj}$                               & \begin{tabular}[c]{@{}l@{}}Element of adjacency matrix denoting connectivity between \\ ONUs $d\in D$ and RNs $b\in B$\end{tabular}                          \\
$x_{cd}^{adj}$                               & \begin{tabular}[c]{@{}l@{}}Element of adjacency matrix denoting connectivity between \\ ONUs $d\in D$ and COs $c\in C$\end{tabular}                          \\
$BW_{dz}$                                    & \begin{tabular}[c]{@{}l@{}}Bandwidth of the optical fiber link between cloudlet at \\ $z\in \{a,b,c\}$ and ONU $d\in D$ in the uplink direction\end{tabular}   \\
$BW_{zd}$                                    & \begin{tabular}[c]{@{}l@{}}Bandwidth of the optical fiber link between cloudlet at \\ $z\in \{a,b,c\}$ and ONU $d\in D$ in the downlink direction\end{tabular} \\
$D_{zd}$                                     & \begin{tabular}[c]{@{}l@{}}Transmission latency between cloudlet at $z\in \{a,b,c\}$ \\ and ONU at $d\in D$\end{tabular}                                       \\
$D_{QoS}$                                    & \begin{tabular}[c]{@{}l@{}}Maximum allowed latency according to desired quality of \\service (QoS)\end{tabular}  \\
$\mu$                                         & Average service rate of each rack in a cloudlet or cloud                                                                                                             \\
$n_{\lambda}$                             & \begin{tabular}[c]{@{}l@{}}Number of wavelengths shared by ONUs to\\ communicate to the corresponding RN cloudlet\end{tabular}   \\
$\lambda_z$                                   & \begin{tabular}[c]{@{}l@{}}Total task arrival rate at cloudlet at $z\in \{a,b,c\}$, from all\\ $d\in D$ ONUs connected to it\end{tabular}                      \\
$\lambda_d$                                  & Task arrival rate from a single ONU at $d\in D$                                                                                                              \\
$\mu_z$                                      & \begin{tabular}[c]{@{}l@{}}Total service rate of a cloudlet at $z\in \{a,b,c\}$ containing \\ $m\in K$ racks\end{tabular}                                      \\
$\Lambda$                                    & \begin{tabular}[c]{@{}l@{}}Mean transmission latency to offload a service request \\to remote cloud\end{tabular}                                            \\
$\sigma_{ul}$                                & \begin{tabular}[c]{@{}l@{}}Average number of bits an ONU $d\in D$ sends  to cloudlet\\ $z\in \{a,b,c\}$ for processing\end{tabular}                            \\
$\sigma_{dl}$                                & \begin{tabular}[c]{@{}l@{}}Average number of bits an ONU $d\in D$ receives from cloudlet\\ $z\in \{a,b,c\}$ after processing\end{tabular}                      \\
$\beta_{dl}$                                 & Background load in the downlink of the considered TDM PON                                                                                                    \\
$\beta_{ul}$                                 & Background load in the uplink of the considered TDM PON                                                                                                      \\ \hline
\end{tabular}
}
\end{table}
\setlength{\textfloatsep}{4pt}

\section{System Model of CCOMPASSION}\label{sec3}
We present the system model and MINLP formulation in this section.  We assume that the CO, RN and ONU locations are already known to us irrespective of the deployment scenario.  We pre-identify some potential field-cloudlet placement locations by using data clustering techniques on the ONU locations.\par
\subsection{Hybrid Cloudlet Network Optimization Parameters}
We assume each cloudlet contains a finite set of processors, e.g., $m\in K=\{1,...,10\}$ racks with service rate of $\mu$ for each rack.  Such a system should be an M/M/$m$ queueing system if it performs the incoming tasks sequentially and total processing delay can be computed using \emph{Erlang-C formula} \cite{MJia}.  Nonetheless, here we model the system differently by assuming that service process of each task is independent of each other and are exponentially distributed with service rate $\mu$, if a task is serviced by a single rack.  We also assume that only one task is performed at a particular instant, homogeneously by all the racks combined, i.e., parallel processing of a task by multiple racks.  The task arrival process to any cloudlet from the corresponding ONUs are also assumed to be independent and Poisson distributed with an arrival rate $\lambda$.  Based on these considerations, we model the cloudlets as M/M/1 queueing systems.  In CCOMPASSION, the number of racks $m$ may vary from cloudlet to cloudlet, depending on number of ONUs they are connected to.  If a cloudlet location (either in the field or RN or OLT) is not chosen, then the total number of racks should also be zero for that particular location.\par
The total incoming service request rate to a cloudlet is calculated by adding all the service requests arriving from all ONUs connected to that cloudlet.  The operational cloudlets may choose either to process the total incoming task request all by itself or offload some fraction to the remote cloud, while meeting the latency constraints.  Offloading incoming tasks to remote cloud is obviously a cost minimizing opportunity for cloudlets, but meeting the latency requirements becomes challenging due to very high end-to-end latency from remote cloud.  When a cloudlet chooses to process the total incoming task request all by itself, the average processing time for any cloudlet in the field, RN or OLT location $z\in \{a,b,c\}$ is expressed as follows \cite{Kleinrock}:
\begin{equation} \label{eq1}
\tau_{z}^{cloudlet}=\frac{1}{(\mu_z-\lambda_z)},\forall z\in \{a,b,c\}
\end{equation}
\par The core cloud possesses huge computational and storage resources and hence can be assumed as an M/M/$\infty$ queuing system.  The total latency for a task when offloaded to cloud can be expressed as follows \cite{MJia}:
\begin{equation} \label{eq2}
\tau^{cloud}=\Lambda+\frac{1}{\mu}
\end{equation}
\par A summary of network optimization parameters with their definitions are presented in Table \ref{table1} for convenience.
\subsection{The Decision Variables}
To formulate the objective function and constraints of our MINLP problem, we choose a set of binary and fractional decision variables as follows. The parameters $x_a$, $x_b$ and $x_c$ are binary variables, indicating the decision whether to install a cloudlet in a field site $a\in A$, in a RN site $b\in B$ and in a CO site $c\in C$, respectively.\par
\begin{equation*}
x_a=\begin{cases}1; & \text{if a cloudlet is installed}\\ & \text{in the field location }  a\in A\\0; & \text{otherwise}\end{cases}
\end{equation*}
\begin{equation*}
x_b=\begin{cases}1; & \text{if a cloudlet is installed}\\ & \text{at RN location }  b\in B\\0; & \text{otherwise}\end{cases}
\end{equation*}
\begin{equation*}
x_c=\begin{cases}1; & \text{if a cloudlet is installed}\\ & \text{at CO location }  c\in C\\0; & \text{otherwise}\end{cases}
\end{equation*}
\par The parameters $n_{am}$, $n_{bm}$ and $n_{cm}$ are binary variables to indicate whether $m$ number of racks are chosen to be placed in cloudlet location $a\in A$, $b\in B$ and $c\in C$, respectively.\par
\begin{equation*}
n_{am}=\begin{cases}1; & \text{if the cloudlet at } a\in A\\ & \text{contains } m \text{ racks}  \\0; & \text{otherwise}\end{cases}
\end{equation*}
\begin{equation*}
n_{bm}=\begin{cases}1; & \text{if the cloudlet at } b\in B\\ & \text{contains } m \text{ racks}  \\0; & \text{otherwise}\end{cases}
\end{equation*}
\begin{equation*}
n_{cm}=\begin{cases}1; & \text{if the cloudlet at } c\in C\\ & \text{contains } m \text{ racks}  \\0; & \text{otherwise}\end{cases}
\end{equation*}
\par We define the following binary variables as the product of some particular combinations of the aforementioned binary variables, i.e., $\overline{x_a}=x_an_{am}$, $\overline{x_b}=x_bn_{bm}$ and $\overline{x_c}=x_cn_{cm}$.\par
The parameters $x_{ad}$, $x_{bd}$ and $x_{cd}$ are binary variables to indicate whether the connectivity between an ONU at $d\in D$ and a cloudlet at $a\in A$, $b\in B$ and $c\in C$ respectively, is created or not.
\begin{equation*}
x_{ad}=\begin{cases}1; & \text{if ONU } d\in D \text{ is connected}\\ & \text{to field cloudlet } a\in A \\0; & \text{otherwise}\end{cases}
\end{equation*}
\begin{equation*}
x_{bd}=\begin{cases}1; & \text{if ONU } d\in D \text{ is connected}\\ & \text{to RN cloudlet } b\in B \\0; & \text{otherwise}\end{cases}
\end{equation*}
\begin{equation*}
x_{cd}=\begin{cases}1; & \text{if ONU } d\in D \text{ is connected}\\ & \text{to CO cloudlet } c\in C \\0; & \text{otherwise}\end{cases}
\end{equation*}
\par The parameters $\varphi_a$, $\varphi_b$ and $\varphi_c$ are fractions lying within the range [0, 1] and indicate the portion of the total incoming task a cloudlet at $a\in A$, $b\in B$ and $c\in C$ choses to process by itself.
\subsection{Objective Function and Constraints}
We formulate the objective function to minimize the overall cloudlet installation expenditures (CAPEX) as follows:
\begin{equation}\label{eq3}
\begin{split}
\min\left\{\sum_{a\in A}\sum_{m\in K}(m\alpha) \overline{x_a}+\sum_{b\in B}\sum_{m\in K}(m\alpha) \overline{x_b}+\sum_{c\in C}\sum_{m\in K}(m\alpha) \overline{x_c} \right.\text{\hspace{10pt}}\\ 
\left.+\sum_{a\in A}\sum_{d\in D}\eta L_{ad} x_{ad}+\sum_{a\in A}\xi_a x_a+\sum_{b\in B}\xi_b x_b+\sum_{c\in C}\xi_c x_c\right\}\text{\hspace{10pt}}
\end{split}
\end{equation}
\par The above expression includes cost components arising from multiple sectors.  The first, second and third terms denote the total cost of racks installed in the field, the RN and the CO cloudlet locations, respectively.  The fourth term denotes the total cost of installing new fiber connections among ONUs and field cloudlets.  Finally, the fifth, sixth and seventh terms denote the additional infrastructure setup costs in the field, the RN and the CO cloudlet locations, respectively.\par
\par Actually the first three terms of the objective function (\ref{eq3}) are quadratic in nature due to $\overline{x_a}=x_an_{am}$, $\overline{x_b}=x_bn_{bm}$ and $\overline{x_c}=x_cn_{cm}$, but we linearize them using three linear inequalities for each of them, as shown in (\ref{eq4})-(\ref{eq6}) \cite{Intprog}.
\begin{eqnarray}
\overline{x_a}\leq n_{am}, \overline{x_a}\leq x_{a}, \overline{x_a}\geq n_{am}+x_a-1, \forall a\in A, m\in K \label{eq4}\\
\overline{x_b}\leq n_{bm}, \overline{x_b}\leq x_{b}, \overline{x_b}\geq n_{bm}+x_b-1, \forall b\in B, m\in K \label{eq5}\\
\overline{x_c}\leq n_{cm}, \overline{x_c}\leq x_{c}, \overline{x_c}\geq n_{cm}+x_c-1, \forall c\in C, m\in K \text{\hspace{3.5pt}}\label{eq6}
\end{eqnarray}
\par The set of constraints (\ref{eq7})-(\ref{eq9}) are the \emph{capacity constraints} and are used to ensure that only one value of $m$ (the number of racks in a particular cloudlet) is chosen.  If a particular cloudlet is not chosen to be activated at either of field, RN or CO location, then no racks are placed at that location.
\begin{gather}
x_a\leq \sum_{m\in K} n_{am} \text{   and  } \sum_{m \in K} n_{am}\leq 1,\forall a\in A, m\in K \label{eq7}\\
x_b\leq \sum_{m\in K} n_{bm} \text{   and  } \sum_{m \in K} n_{bm}\leq 1,\forall b\in B, m\in K \label{eq8}\\
x_c\leq \sum_{m\in K} n_{cm} \text{   and  } \sum_{m \in K} n_{cm}\leq 1,\forall c\in C, m\in K \label{eq9}
\end{gather}
\par Constraints described next are \emph{connectivity constraints}.  Constraint (\ref{eq10}) denotes that any ONU can be chosen to be connected to a field cloudlet, only if their Euclidean separation is  $\leq L_{max}$. Constraint (\ref{eq11}) denotes that a field cloudlet will be activated of there exists at least one connected ONU.
\begin{gather}
x_{ad}\leq \max\left[0,\left(\frac{L_{ad}-L_{max}}{|L_{ad}-L_{max}|}\right)\right],\forall a\in A,d\in D \label{eq10}\\
x_a\geq x_{ad}, \forall a\in A,d\in D \label{eq11}
\end{gather}
\par In a TDM-PON, any arbitrary ONU cannot be connected to any RN cloudlet, rather only those ONUs can be connected which already has an existing fiber connection to a particular RN and the same condition applies for connectivity between ONU and CO as well.  These conditions are enforced by constraints (\ref{eq12})-(\ref{eq13}).  Furthermore, constraints (\ref{eq14})-(\ref{eq15}) imply that a cloudlet in the RN and in the CO can be activated if there is at least one ONU connected to it, respectively.
\begin{gather}
x_{bd}\leq x_{bd}^{adj},\forall b\in B,d\in D \label{eq12}\\
x_{cd}\leq x_{cd}^{adj},\forall c\in C,d\in D \label{eq13}\\
x_b\geq x_{bd}, \forall b\in B,d\in D \label{eq14}\\
x_c\geq x_{cd}, \forall c\in C,d\in D \label{eq15}
\end{gather}
\par The constraint in (\ref{eq16}) ensures that every ONU is connected to one and only one cloudlet.
\begin{equation}
\sum_{a\in A} x_{ad}+\sum_{b\in B} x_{bd}+\sum_{c\in C} x_{cd}=1,\forall d\in D \label{eq16}
\end{equation}
\par Finally, the most important constraint, i.e., the \emph{QoS latency constraint} that provides the upper limit of total allowed latency for each ONU, is described in (\ref{eq17}).
\begin{flalign}\label{eq17}
x_{ad}\left[\varphi_a\left\{\frac{1}{\mu_a-\lambda_a\varphi_a}+D_{ad}+\frac{\sigma_{ul}}{BW_{da}}+\frac{\sigma_{dl}}{BW_{ad}}\right\}\right.\nonumber\text{\hspace{60pt}}\\
\left.+(1-\varphi_a)\left(\Lambda+\frac{1}{\mu}\right)\right]\nonumber\\
+x_{bd}\left[\varphi_b\left\{\frac{1}{\mu_b-\lambda_b\varphi_b}+D_{bd}+\frac{\sigma_{ul}\sum_{d\in D}x_{bd}}{n_{\lambda}BW_{db}}+\frac{\sigma_{dl}\sum_{d\in D}x_{bd}}{n_{\lambda}BW_{bd}}\right\}\right.\nonumber\\
\left.+(1-\varphi_b)\left(\Lambda+\frac{1}{\mu}\right)\right]\nonumber\\
+x_{cd}\left[\varphi_c\left\{\frac{1}{\mu_c-\lambda_c\varphi_c}+D_{cd}+\frac{\sigma_{ul}\sum_{d\in D}x_{cd}}{BW_{dc}-\beta_{ul}}+\frac{\sigma_{dl}\sum_{d\in D}x_{cd}}{BW_{cd}-\beta_{dl}}\right\}\right.\nonumber\\
\left.+(1-\varphi_c)\left(\Lambda+\frac{1}{\mu}\right)\right]\leq D_{QoS},\forall a\in A,b\in B, c\in C,d\in D
\end{flalign}
\begin{figure*}[t!]
\centering
\includegraphics[width=\textwidth,height=5cm]{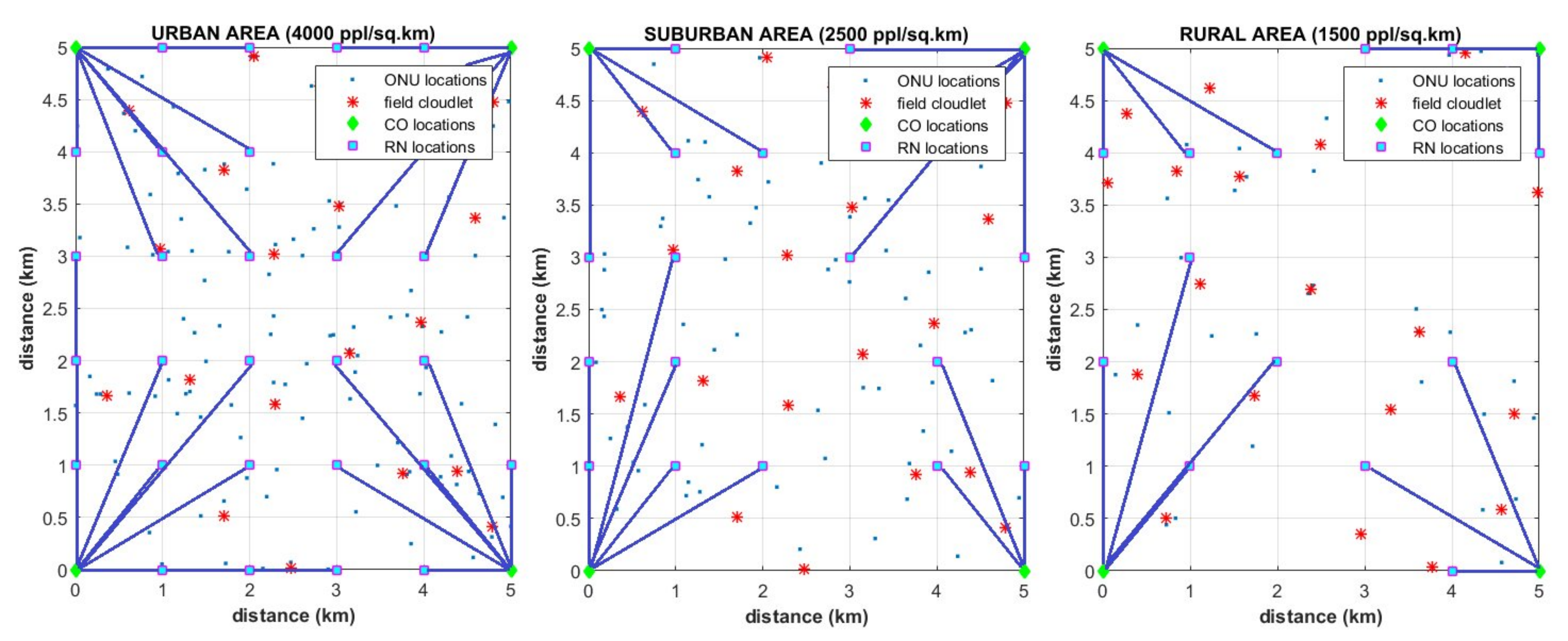}
\caption{Randomly generated urban, suburban and rural areas for cloudlet deployment with existing TDM-PON network infrastructure with 1:4 split ratio (The blue links from CO to RN locations represent the feeder fibers. Distribution fibers from RN to each ONU are not shown).}
\label{dataset}
\end{figure*}
\setlength{\textfloatsep}{4pt}
\begin{figure*}[h!]
\centering
\includegraphics[width=\textwidth,height=5cm]{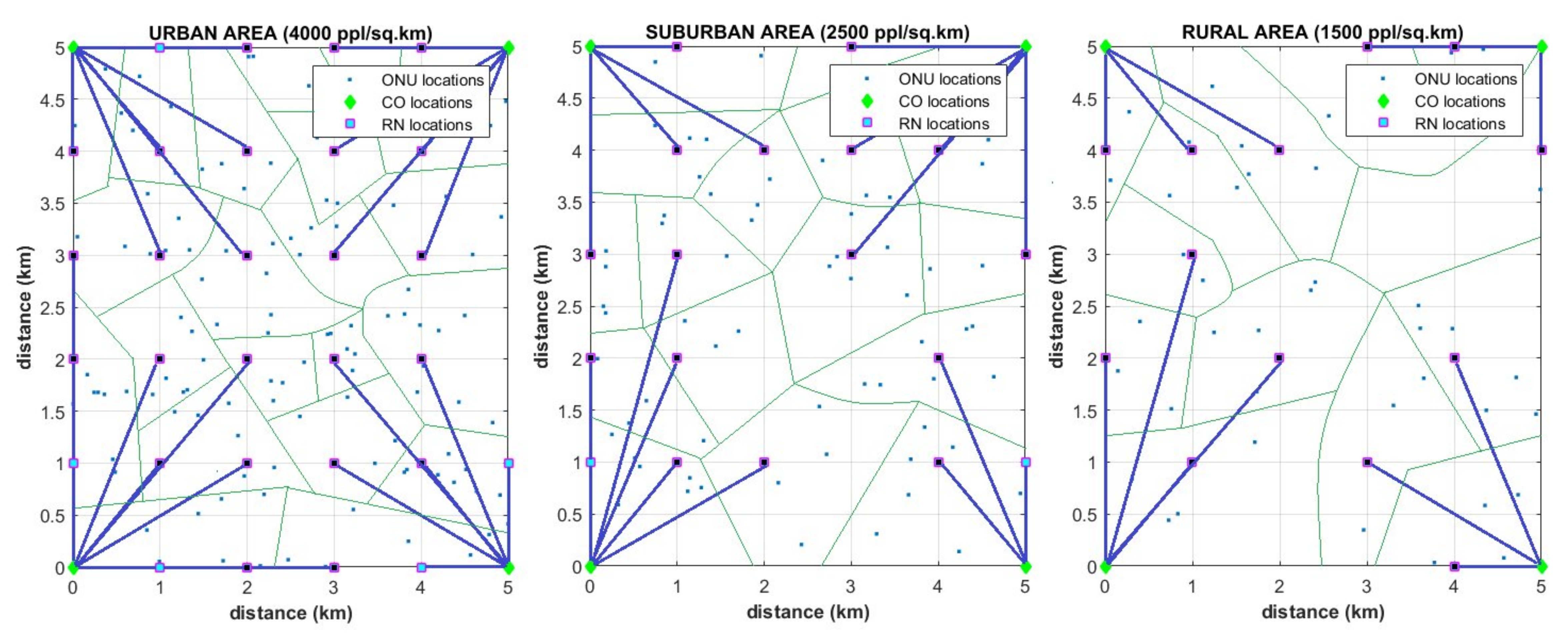}
\caption{Optimal cloudlet placement locations over the considered urban, suburban and rural areas for 1 ms latency constraint using CCOMPASSION (cloudlets are installed at darkened face RNs and all COs).}
\label{opt_dataset}
\end{figure*}
\setlength{\textfloatsep}{4pt}
\par Note that, the total task arrival rate to any cloudlet, depending on connectivity among ONUs and cloudlets, and the total service rate of any cloudlet is computed as follows:
\begin{equation}
\lambda_z=\sum_{d\in D}x_{zd}\lambda_d,\text{   and   }\mu_z=\sum_{m\in K}mn_{zm}\mu,\forall z\in \{a,b,c\} \label{eq18}
\end{equation}
\par The constraint (\ref{eq17}) consists of the weighted sum of total latencies when the incoming tasks are either executed in the field or RN or CO cloudlets.  However, due to (\ref{eq10}) each ONU can connect to only one cloudlet.  Each of these weighted latency terms consists of four components.  The first component denotes the processing time, the second component denotes transmission time, the third component denotes the time to offload total bits for the task by ONU to cloudlet and the fourth component denotes the time to receive total bits post-processing by ONU from cloudlet.  The third and fourth components are evaluated differently from each other, depending on the respective data transmission schemes from ONUs to cloudlets in the field, at RN or at CO.

\section{CCOMPASSION Framework Evaluation}\label{sec4}
We implemented the MINLP model described in Section \ref{sec3} with the \emph{A Modeling Language for Mathematical Programming} (AMPL) platform and solved the problem using the open-source solver COUENNE package for MINLPs \cite{Couenne}.  However, we understand that MINLP problems are usually \emph{NP-Hard} in nature and the COUENNE package uses some variations of \emph{Branch-and-Bound algorithm} that may take several hours to days to compute the optimal solution, depending on volume of dataset and available computational capacity \cite{algorithms}.  Hence, a time-scalable heuristic algorithm specific to CCOMPASSION needs to be developed, but during the network planning stage, network operators can tolerate the required computational time.\par
\begin{figure*}[t!]
\centering
\includegraphics[width=\textwidth,height=5cm]{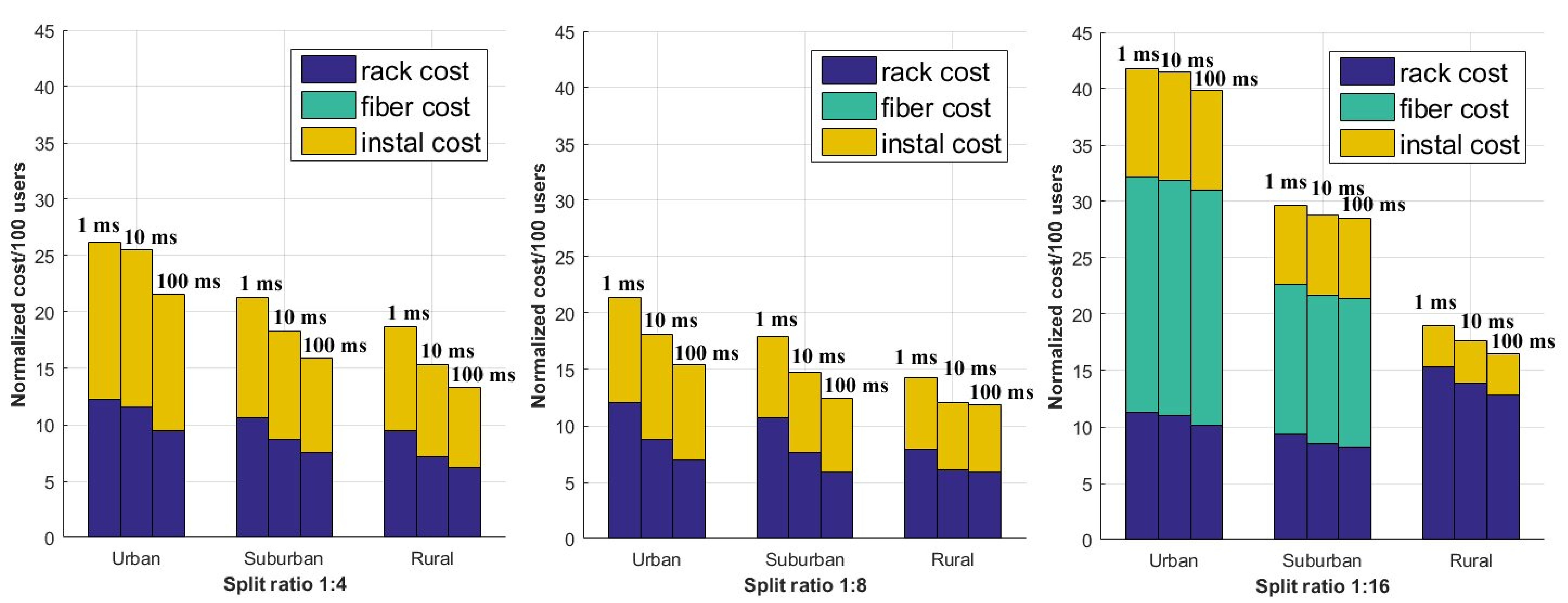}
\caption{Normalized cost/100 users in urban, suburban and rural scenarios against $D_{QoS}$ = 1 ms, 10 ms, and 100 ms.}
\label{norm_cost}
\end{figure*}
\setlength{\textfloatsep}{4pt}
\begin{figure*}[h!]
\centering
\includegraphics[width=\textwidth,height=5cm]{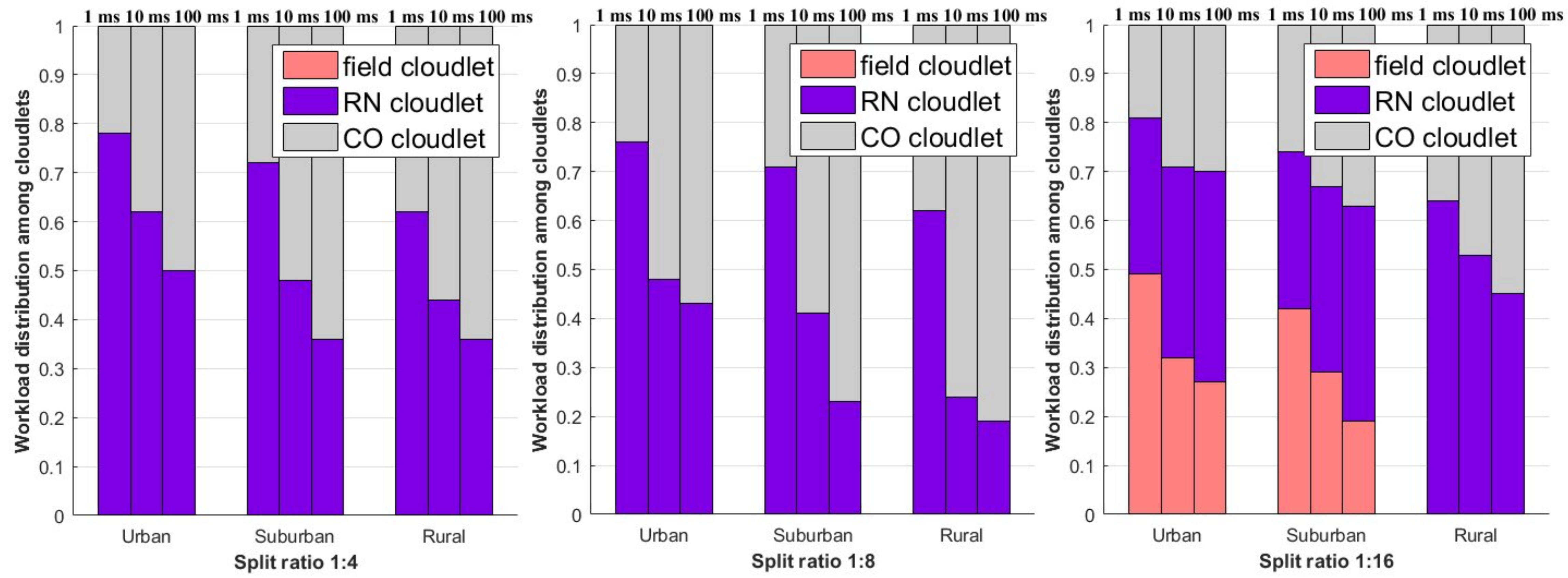}
\caption{Distribution of workload among cloudlets in the field, RN and CO locations in urban, suburban and rural scenarios against $D_{QoS}$ = 1 ms, 10 ms, and 100 ms.}
\label{work_dist}
\end{figure*}
\setlength{\textfloatsep}{4pt}
\subsection{Random Dataset Generation}
To test the CCOMPASSION framework, we used stochastically-generated hypothetical 5 km x 5 km Australian city/towns to represent urban, suburban and rural deployment scenarios.  Following the guidelines provided in \cite{Australia}, we consider the population densities of 4000, 2500 and 1500 per square kilometre for urban, suburban and rural areas, respectively.  We used the \emph{Poisson-point process} to distribute the total population across the considered area to identify the ONU locations and followed by used \emph{k-means clustering algorithm} to identify potential field cloudlet placement locations \cite{Matthews}.  The value of $k$ needs to be sufficiently large enough to maintain the feasibility of the optimal solution and we arbitrarily choose it as 20.  The ONUs are supported by the standard 10G(E)-PON infrastructure with 1:$N$, $N\in\{4,8,16\}$ split ratio and 10 Gbps datarate in both downlink and uplink.  Fig. \ref{dataset} presents the considered datasets.  The small blue dots represent the ONU locations, the red asterics represent the potential field cloudlet locations, the cyan squares represent the RN locations and the green diamonds represent the CO locations (corner points of the 5 km x 5 km block).  The solid blue lines indicate the connectivity amongst the RNs and COs. Note that feeder fiber and distribution fiber connectivity between RNs and ONUs is not shown.\par
\subsection{Values of Network Parameters and Cost components}
The normalized costs of installing new point-to-point fiber per kilometer ($\eta$) are 50, 35 and 20 for urban, suburban and rural, respectively \cite{Machuka}.  The datarate of these new point-to-point fiber links are considered to be $BW_{ad}$ = $BW_{da}$ = 1 Gbps and the maximum allowed distance among field cloudlets and ONUs is $L_{max}$ = 4 km, such that feasibility of optimal solution is maintained.  We consider that $n_{\lambda}$ = 1 wavelength (with a maximum 10 Gbps datarate is shared among ONUs) to communicate with the cloudlet at their corresponding RN location and hence $BW_{bd}$ = $BW_{db}$ = 10 Gbps.  As ONUs are allowed to use the default uplink and downlink channels to communicate with cloudlets at their corresponding CO, therefore $BW_{cd}$ = $BW_{dc}$ = 10 Gbps.  We consider the approximate downlink and uplink background loads as $\beta_{dl}$ = 7 Gbps and $\beta_{ul}$ = 5 Gbps, respectively\cite{Machuka}.  The normalized cost of installing one rack is considered as $\alpha$ = 1 \cite{Lenovo}.  We also consider the normalized costs of installing new cloudlet infrastructure at field and RN locations equal, i.e., $\xi_a$ = $\xi_b$ = 4 and less at CO locations, i.e., $\xi_c$ = 2 \cite{Lenovo}.  We assumed that each ONU can serve a maximum 1000 users and hence the maximum task arrival rate from each ONU $d\in D$ is $\lambda_d$ = 1000 VMs/sec.  In \cite{Ha}, the authors have outlined a summary of average request and response sizes of some commonly used applications and thus we consider $\sigma_{ul}$ = 1 MB and $\sigma_{dl}$ = 1 KB.  The service rate of each rack in cloud or cloudlet ($\mu$) is 2500 VMs/sec \cite{Ceselli}.  The end-to-end latency to offload a task to cloud server ($\Lambda$) is relatively large and considered to be 0.8 sec \cite{MJia}.\par
\begin{figure*}[t!]
\centering
\includegraphics[width=\textwidth,height=5cm]{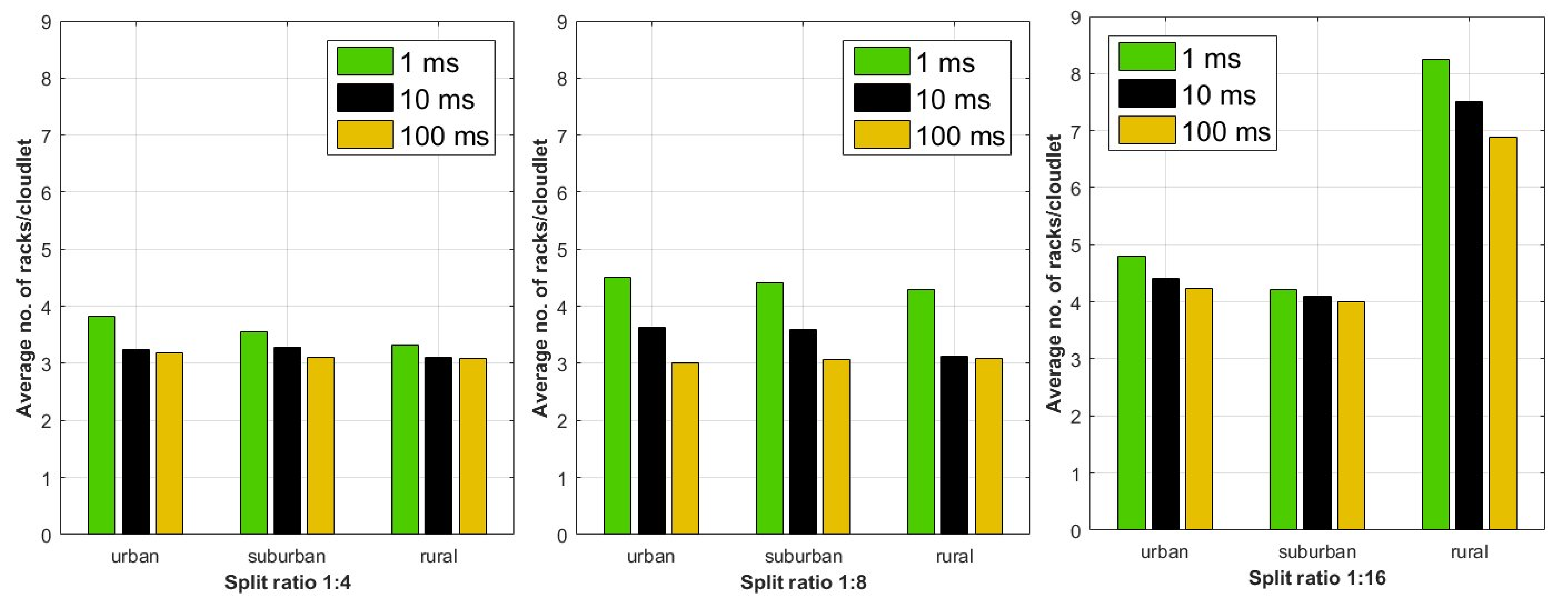}
\caption{Average number of racks/cloudlet in urban, suburban and rural scenarios against $D_{QoS}$ = 1 ms, 10 ms, and 100 ms.}
\label{avg_racks}
\end{figure*}
\setlength{\textfloatsep}{4pt}
\begin{figure*}[h!]
\centering
\includegraphics[width=\textwidth,height=5cm]{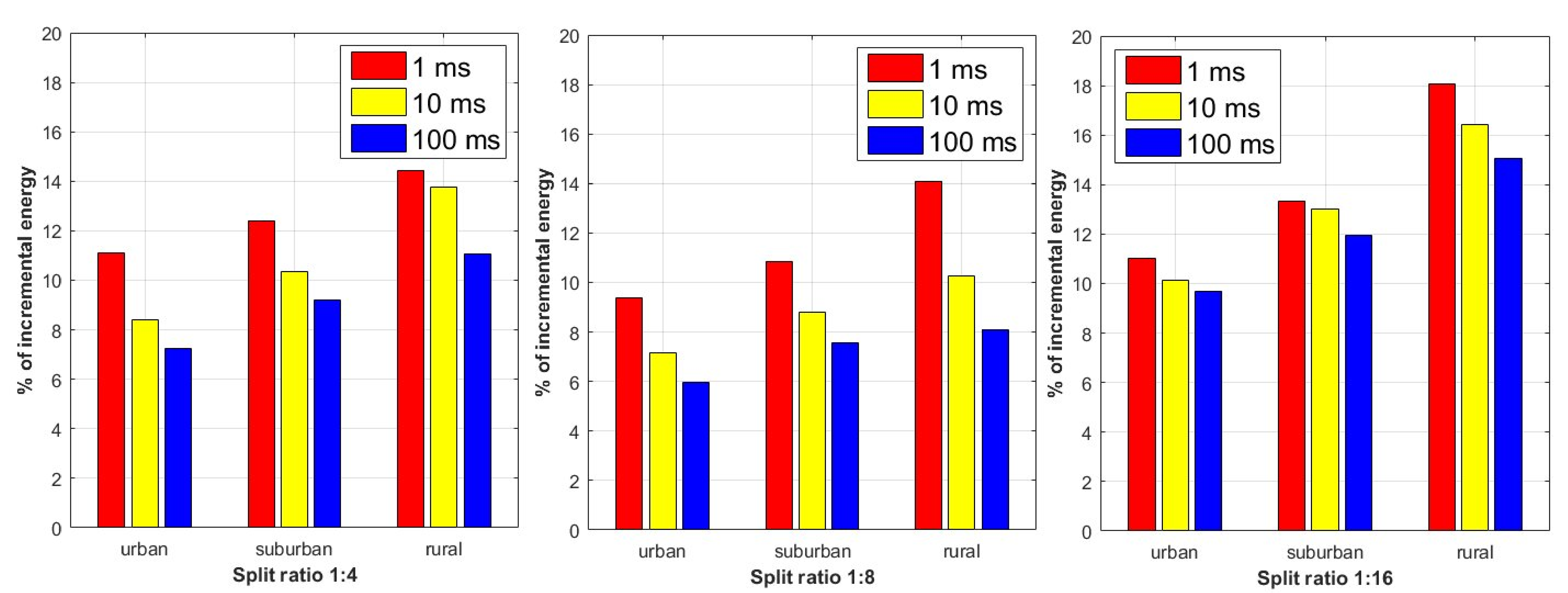}
\caption{Incremental energy consumption over existing TDM-PON in urban, suburban and rural scenarios against $D_{QoS}$ = 1 ms, 10 ms, and 100 ms.}
\label{energy_excess}
\end{figure*}
\setlength{\textfloatsep}{4pt}
\subsection{Results and Discussions}
In this section, we discuss different aspects of the optimal solutions obtained through evaluating CCOMPASSION framework on the urban, suburban and rural deployment scenarios against target latency 1 ms, 10 ms and 100 ms.  Fig. \ref{opt_dataset} shows the optimal cloudlet placement locations in urban, suburban and rural scenarios with 1:4 split ratio against $D_{QoS}$ = 1 ms latency target.  For this case, field cloudlets are not chosen, rather only some of the RN locations (darkened) and all CO locations are chosen.  The green borders indicate approximate coverage area of each of the cloudlets installed.  Fig. \ref{norm_cost} shows the cost of cloudlet installation normalized to 100 users for the all three $D_{QoS}$ values of 1 ms, 10 ms and 100 ms.  The total cost consists of three components, e.g., cost of racks, cost of newly installed fiber and cost of new infrastructure setup.  The bar graphs imply that the cost of installing cloudlets in the urban scenario is the highest and the least in rural scenario with all split ratios.  This is primarily due to the fact that more cloudlets are required in the urban and suburban scenarios compared to rural scenario to meet latency requirements.  With a split ratio 1:16, a higher number of ONUs must share the same wavelength as compared to 1:4 and 1:8 split ratio scenarios and hence the per user average bandwidth is lower.  As such, for urban and suburban scenarios with a 1:16 split ratio, cloudlets are required to be installed in the field, RN and CO locations, but for the rest of the cases only RN and CO cloudlets are required to be installed.  Hence, the normalized cost of cloudlet deployment in urban and suburban scenarios with a 1:16 split ratio is relatively higher than other cases.\par
Fig. \ref{work_dist} shows the distribution of workload amongst the cloudlets installed in the field, RN or CO for urban, suburban and rural scenarios with 1:$N$, $N\in\{4,8,16\}$ split ratios, for the all three $D_{QoS}$ values of 1 ms, 10 ms and 100 ms.  It is very interesting to observe that for all deployment scenarios, when $D_{QoS}$ = 1 ms, a majority of the computational tasks are performed by the field and RN cloudlets.  However, as the QoS latency requirement is relaxed, i.e., with $D_{QoS}$ = 10 ms and 100 ms, a higher percentage of workload is handled by the CO cloudlets.  Therefore, when the QoS latency requirement is stringent, expensive cloudlets are required to be installed in the field or RN depending on the deployment scenario. In contrast, with a less stringent QoS latency requirement, installation of more economical CO cloudlets could be preferred.\par
The average number of racks required per cloudlet is shown in Fig. \ref{avg_racks} for the urban, suburban and rural scenarios with $1:N$, $N\in\{4,8,16\}$ split ratios against $D_{QoS}$ values of 1 ms, 10 ms and 100 ms.  When $D_{QoS}$ = 1 ms, it can be interpreted from (\ref{eq1}) that a higher number of racks are required, whereas when $D_{QoS}$ = 10 ms or 100 ms, a relatively lower number of racks are required to process the same number of VMs.  This fact is also very much aligned with our observations from the results summarized in Fig. \ref{avg_racks}.  We also note that the average number of racks per cloudlet in all scenarios with split ratio 1:8 is higher than with split ratio 1:4, because of the higher number of available RN locations with the 1:4 split ratio.  Ideally, in a cost minimization framework a minimum number of cloudlets should be installed with an optimal number of racks.  However, with a split ratio of 1:16, a few field cloudlets have to be installed in the urban and suburban scenarios.  Thus in this case, the average number of racks per cloudlets in urban and suburban scenarios is lower as compared to the average number of racks in the rural scenario.\par
Installation of cloudlets over an existing 10G(E)-PON infrastructure increases the overall network energy budget.  In Fig. \ref{energy_excess}, we present this percentage of increase in energy consumption for the urban, suburban and rural scenarios with 1:$N$, $N\in\{4,8,16\}$ split ratios against $D_{QoS}$ values of 1 ms, 10 ms and 100 ms.  We considered the energy consumption of different elements of a 10G(E)-PON viz., OLT, ONU, wireless links etc. according to \cite{Machuka} and the energy consumption of cloudlets according to the technical specifications of Lenovo ThinkStation P900 \cite{Lenovo}.  Note that, albeit the urban and suburban scenarios have more cloudlets installed, the percentage of incremental energy budget is the highest for rural scenario against all $D_{QoS}$ values of 1 ms, 10 ms and 100 ms.  This is due to the small overall energy budget of rural deployment as compared to the urban and suburban deployment.  Thus installing a single cloudlet in the rural scenario increases the energy budget more than urban and suburban scenarios.  However, for all considered 5 km x 5 km urban, suburban and rural scenarios with $1:N$, $N\in\{4,8,16\}$ split ratios, the percentage of incremental energy budget in the presence of cloudlets are less than 20\%.\par

\section{Summary}\label{sec5}
The idea of deploying distributed cloudlet facility solves the majority of the network bottleneck issues in the converged mobile computing and cloud computing environment and hence provisioning of many computation-intensive and interaction-intensive applications have become possible.  Meeting the strict latency requirements of these applications has been a primary challenge due to latency in accessing the cloud services and optical access networks appeared as a promising solution in this aspect due to its low-cost per bit and huge bandwidth support.  In this paper, we have proposed a hybrid cloudlet placement framework CCOMPASSION over existing TDM-PON infrastructures and have developed a MINLP model to identify cost optimal cloudlet placement locations.  We tested this framework over hypothetical urban, suburban and rural deployment scenarios with 1:$N$, $N\in\{4,8,16\}$ split ratios to achieve target latency values 1 ms, 10 ms and 100 ms.  Through the evaluation of this framework, various seminal insights on the optimal deployment strategies of cloudlets in urban, suburban and rural deployment scenarios were provided.



\bibliographystyle{IEEEtran}

\bibliography{IEEEabrv,ref_infocom}

\end{document}